\documentstyle[12pt]{article}

\catcode`\@=11

\def\@fnsymbol#1{\ifcase#1\or * \or \dagger\or \ddagger\or
\mathchar "278\or \mathchar "27B\or \|\or **\or \dagger\dagger
\or \ddagger\ddagger \else\@ctrerr\fi\relax}%

\def\title{%
\vspace{0.5cm}\vspace{4ex}
\bgroup
\obeylines
\large\boldmath \bf\begin{center}
}
\def\endtitle{\end{center}\vskip1sp\egroup}
\def\author#1{\begingroup\center #1 \endcenter\endgroup}

\def\keywords#1{\par\noindent{\bf KEYWORDS}: #1}
\def\addcontentsline#1#2#3{\relax}
\def\fnum@figure{Figure \thefigure}
\def\fnum@table{Table \thetable}
\newcounter{figcaption}
\def\thefigcaption{\arabic{figcaption}}
\def\fnum@figcaption{{\bf Fig. \thefigcaption :}}
\def\figcaption{%
 \par \vskip 10pt
\list{\fnum@figcaption}
{\leftmargin 5em \labelwidth\leftmargin\advance\labelwidth-\labelsep
\def\makelabel##1{##1\hfil} \usecounter{figcaption}}%
}

\def\cite{\@ifnextchar [{\@tempswatrue\@citex}{\@tempswafalse\@citex[]}}
\def\@citex[#1]#2{\if@filesw\immediate\write\@auxout{\string\citation{#2}}\fi
\def\@citea{}\@cite{\@for\@citeb:=#2\do
{\if-\@citeb \mbox{-}\def\@citea{}
\else
\@citea\def\@citea{,\penalty\@m}\@ifundefined
{b@\@citeb}{{\bf ?}\@warning
{Citation `\@citeb' on page \thepage \space undefined}}%
\hbox{\csname b@\@citeb\endcsname}
\fi}}{#1}}
\newfont{\scrptrm}{cmr8}
\def\@cite#1#2{${}^{\scrptrm {#1\if@tempswa , #2\fi})}$}

\def\rcite{\@ifnextchar [{\@tempswatrue\@rcitex}{\@tempswafalse\@rcitex[]}}
\def\@rcitex[#1]#2{\if@filesw\immediate\write\@auxout{\string\citation{#2}}\fi
\def\@rcitea{}\@rcite{\@for\@rciteb:=#2\do
{\if-\@rciteb \mbox{-}\def\@rcitea{}%
\else
\@rcitea\def\@rcitea{,\penalty\@m}\@ifundefined
{b@\@rciteb}{{\bf ?}\@warning
{Citation `\@rciteb' on page \thepage \space undefined}}%
\hbox{\csname b@\@rciteb\endcsname}%
\fi}}{#1}}
\def\@rcite#1#2{{#1\if@tempswa , #2\fi}}
\def\refcite{\@ifnextchar [{\@tempswatrue\@refcitex}
{\@tempswafalse\@refcitex[]}}
\def\@refcitex[#1]#2{
\if@filesw\immediate\write\@auxout{\string\citation{#2}}\fi
\def\@citea{}\@refcite{\@for\@citeb:=#2\do
{\if-\@citeb -\def\@citea{}
\else
\@citea\def\@citea{,\penalty\@m}\@ifundefined
{b@\@citeb}{{\bf ?}\@warning
{Citation `\@citeb' on page \thepage \space undefined}}%
\hbox{\csname b@\@citeb\endcsname}
\fi}}{#1}}
\def\@refcite#1#2{[{#1\if@tempswa , #2\fi}]}

\newcommand{\etal}{{\em et al.}}

\newcommand{\rmc}{{\rm c}}
\newcommand{\rmd}{{\rm d}}
\newcommand{\rme}{{\rm e}}
\newcommand{\rmi}{{\rm i}}

\renewcommand{\Re}{{\cal R}\!{\sl e}\,}
\renewcommand{\Im}{{\cal I}\!{\sl m}\,}

\newcommand{\Ham}{{\cal H}}
\newcommand{\dags}{^\dagger}

\def\braket#1{\left\langle#1\right\rangle}

\def\simle{\mathrel{\mathpalette\@versim<}}   % < over \sim
\def\simge{\mathrel{\mathpalette\@versim>}}   % > over \sim
\def\@versim#1#2{\lower2.5pt\vbox{\baselineskip0pt \lineskip-.5pt
   \ialign{$\m@th#1\hfil##\hfil$\crcr#2\crcr\sim\crcr}}}

\newcommand{\bequ}{ \begin{equation} }
\newcommand{\eequ}{ \end{equation} }
\newcommand{\barr}{ \begin{array} }
\newcommand{\earr}{ \end{array} }
\newcommand{\beqarr}{ \begin{eqnarray} }
\newcommand{\eeqarr}{ \end{eqnarray} }

\catcode`\@=12

\def\MFreq{\rmi \omega_n}

\def\LSMO{La$_{1-x}$Sr$_x$MnO$_3$}

\begin{document}

\begin{title}
Magnetoresistance of the  Double-Exchange Model
in Infinite Dimension
\end{title}

\author{Nobuo {\sc Furukawa}}
\begin{instit}
  Institute for Solid State Physics,\\
  University of Tokyo, Roppongi 7-22-1,\\
  Minato-ku, Tokyo 106
\end{instit}

\begin{abstract}
Double-exchange model in infinite dimension
 is studied as the strong Hund's coupling limit $J\to\infty$ of
the Kondo lattice model.
Several quantities such as Green's function and the d.c.\ conductivity
are calculated in analytical forms.
Magnetoresistance in lightly doped {\LSMO}
is reproduced very well.
\end{abstract}

%\vspace{1cm}
\vfil


\pagebreak

Physics of the strongly correlated electron systems is one of the
most challenging fields in the material science. Especially, after the
discovery of the high-$T_\rmc$ superconducting oxides,
many related compounds of $3d$ transition-metal oxides have been
revisited in an extensive way.
One of such materials is the
manganese oxides with the perovskite-type structure
 $R_{1-x}A_x$MnO$_3$,
where $R$ and $A$ denote  trivalent rare-earth ions and  divalent
alkaline-earth ions, respectively.
The most eminent feature in this family of materials
is  the giant magnetoresistance (MR)
with negative sign.
In the experiment of {\LSMO},\cite{Tokura94,Urushibara9x}\
the negative MR value is  scaled at the small magnetization region as
\bequ
  \frac{\rho(0)-\rho(M_{\rm tot})}{\rho(0)}
   =   C_{\rm exp} \left(\frac{M_{\rm tot}}{M_{\rm s}}\right)^2,
    \label{expUnivFit}
\eequ
where the coefficient $C_{\rm exp}$ is nearly temperature-independent.
Here, $M_{\rm tot}$ and $M_{\rm s} = 4 \mu_{\rm B}$ are
the total magnetization and the nominal saturation magnetization,
respectively, and  $\rho$ is the resistivity.

{}From the theoretical point of view, the author has calculated the
Kondo lattice model with Hund's ferromagnetic coupling
in the infinite-dimensional limit $D=\infty$ and
the infinite high-spin limit
$S=\infty$.
Results  with respect to MR\cite{Tokura94,Furukawa94}
as well as optical conductivity\cite{Furukawa95a} and
magnetic transition temperature\cite{Furukawa95bx}
 are in good agreement with experimental data of {\LSMO}.
The Hamiltonian is given by
\bequ
  \Ham =
  - \sum_{ij,\sigma} t_{ij}
        \left(  c_{i\sigma}\dags c_{j\sigma} + h.c. \right)
    -J \sum_i \vec \sigma_i \cdot \vec m_i,
    \label{HamSinfty}
\eequ
where $ \vec m_i = (m_i{}^x, m_i{}^y, m_i{}^z)$ and
$|\vec m|^2 = 1$. %, while $\vec \sigma$ are Pauli matrices.
Here, itinerant fermions and localized spins represent
electrons in $e_{\rm g}$ orbitals and $t_{2\rm g}$ orbitals of Mn ions,
respectively.
Fermion concentration in {\LSMO} is nominally considered to be
$n=1-x$.
{}From the recent band calculation,\cite{Hamada94x}\
the bandwidth of the itinerant electron is estimated to be
$W \sim 1{\rm eV}$.
Since  Hund's coupling is considered to be
larger than the bandwidth,  the system is regarded to be in
the strong coupling region.
In the limit $J/W=\infty$,  the model is
identical to the double-exchange model.\cite{Kubo72,Zener51}\ \

Although the experimental results have been well explained by the
above model at finite $J/W$,
it is still worthwhile to examine the properties at $J/W\to\infty$:
As we will show below, several quantities such as
Green's functions and d.c.\ conductivity are calculated analytically
in a simple form,
which helps us to obtain physical intuition on  properties of the model
in the strong coupling region.
Analytical expressions are also convenient for the data analysis
in experimental and numerical researches.
In this paper,
we study the double-exchange model in  $D=\infty$
as the strong coupling limit $J/W\to\infty$
 of the Kondo lattice model with $S=\infty$.

Infinite-dimensional system is investigated using the
effective single-site approach.
Green's function is obtained exactly as\cite{Furukawa94}
\bequ
  G =  \left\langle \left (
	G_0{}^{-1} + J \vec m \vec \sigma
	\right)^{-1} \right \rangle,
    \label{fmlG}
\eequ
where the thermal average $\braket{\cdots}$ is taken with respect to the
orientation of the local spin $\vec m$.
Self-energy is given by $\Sigma = G_0{}^{-1} - G^{-1} $.
The Weiss field $ G_0$ should be
determined self-consistently from
\beqarr
  G_0{}^{-1} &=& G_{\rm loc}{}^{-1} + \Sigma,
			\\
  G_{\rm loc} &=& \int \rmd \varepsilon N_0(\varepsilon)
      [\MFreq - (\varepsilon-\mu) - \Sigma ]^{-1}.
\eeqarr
Here we consider the case of the Lorentzian density of states
with the bandwidth $W$,
$ N_0(\varepsilon) = (1/{\pi}) \cdot  ({W}/({\varepsilon^2 + W^2}))$.

At $J \gg W$, the spectral weight splits into two sub-bands at
$\omega \sim \pm J$.\cite{Furukawa95a}\ \
Since we restrict ourselves to the hole doped case $n < 1$, we may
treat the lower sub-band only.
Therefore, we describe the chemical potential as
$ \mu = -J + \delta\mu$ where $ \delta\mu = O(W)$.
{}From the self-consistency equation, $G_0$ is given by
\beqarr
  G_0 (\omega+\rmi\eta)
  &=& (\Omega - J + \rmi W)^{-1}.
    \label{defG0}
\eeqarr
Here, $\Omega \equiv \omega + \delta\mu = O(W)$
is the energy which is measured
from the center of the lower sub-band $-J$.

Magnetic field in the $z$ direction
is applied to the localized spins in the paramagnetic phase,
and  the induced magnetization is expressed as $ M = \braket{m_z}$.
{}From eqs.~(\ref{fmlG}) and (\ref{defG0}),
Green's function is given by
\beqarr
  G_\sigma(\omega+\rmi\eta)
   &=& \frac{ (\Omega - J + \rmi W) - JM\sigma}
	    { (\Omega - J + \rmi W)^2 - J^2}
		\nonumber \\
   &=&  \frac{1+M\sigma}2\frac1{\Omega + \rmi W} + O(1/J).
    \label{defG}
\eeqarr
At $J/W\to\infty$, the spectral weight is calculated as
\beqarr
  A_\sigma(\omega) &=& -\Im G_\sigma(\omega+\rmi\eta) / \pi
		\nonumber\\
   & =& \frac{1+M\sigma}{2}\cdot\frac1{\pi}\frac{W}{\Omega^2+W^2}.
  \label{defQPdos}
\eeqarr
We see that the center of the spectral weight
is indeed shifted to $-J$. The amplitude of $A_\sigma$ is proportional to the
population of the local spin parallel to $\sigma$, which indicates
that the electronic states
 that are anti-parallel to the local spin are projected out.

The self-energy is calculated from eqs.~(\ref{defG0})
and (\ref{defG}) as
\beqarr
  \Sigma_\sigma(\omega+\rmi\eta)
  &=& -J - \frac{1-M\sigma}{1+M\sigma}(\Omega+\rmi W).
  \label{defSigma}
\eeqarr
Then,  eq.~(\ref{defSigma}) gives
 $\Re \Sigma \sim -J$, so the shift in $\mu$ is
self-consistently justified again.
We also see $\Im \Sigma = -W$ at $M=0$,
which means that the quasi-particle excitation is very incoherent;
the lifetime of a  quasi-particle is comparable with the
length of time that an electron transfers site to site.
%Therefore, it is justified to take the $D=\infty$ limit which is
%essentially a single-site treatment.
In the strong coupling region, quasi-particles
loose their coherence in a macroscopic scale due to the
inelastic scattering by thermally fluctuating spins.
The divergence of $\Im\Sigma_\sigma$ is observed at $M\sigma\to -1$,
in accordance with the diminishment of $A_\sigma$,
since the propagation of the quasi-particle with spin anti-parallel to
the magnetization is energetically forbidden at $\omega \sim -J$.

Now, we calculate the MR value.
At finite $J/W$, the MR value has been
 obtained from the Kubo formula as
\bequ
%  [ {\rho(0) - \rho(M)}] / {\rho(0)} = C M^2
  \frac{\rho(0) - \rho(M)}{\rho(0)} = C M^2
   \label{defMRgeneric}
\eequ
at small magnetization regime, where $C$ is
 a function of $x$ and $J/W$.\cite{Furukawa94}\ \
We have $C=1$ at $J\ll W$, and
 $C$ increases  as $J/W$ is increased.
Here we study  the  limit $J/W\to\infty$.
Conductivity in infinite dimension is
calculated from the formula%\cite{Moller92,Pruschke93a}
\beqarr
  \sigma_{\rm dc}  &=& \sigma_0 W^2
  \sum_\sigma \int N_0(\epsilon) \rmd \epsilon
\int \rmd \omega
   \left(-\frac{ \partial f }{\partial \omega} \right)
      A_\sigma{}^2(\epsilon,\omega) ,
   \label{defDCcond}
	\\
  A_\sigma(\epsilon,\omega) &=& -\frac1\pi\Im
  \frac1{\omega-(\epsilon-\mu) - \Sigma_\sigma(\omega+\rmi\eta)},
     \label{defSPfun}
\eeqarr
where $f$ is the Fermi distribution
function.
The constant $\sigma_0$ gives the unit of conductivity.
Hereafter we restrict ourselves to the low temperature regime $T \ll W$.
{}From eqs.~(\ref{defSigma}), (\ref{defDCcond}) and (\ref{defSPfun}),
we have
\beqarr
 \sigma_{\rm dc}(M) &=&  \sigma_{\rm dc}(0) \times r(M),
     \label{relDCcond} \\
%
%    r(M) &=& [{ 1 + 3(1+B) M^2 + B M^4} ] / ({1-M^2}),
    r(M) &=& \frac{ 1 + 3(1+B) M^2 + B M^4}{1-M^2},
	\label{defrofM}
\eeqarr
where $\sigma_{\rm dc}(0)$ and $B$ are functions of the
hole concentration $x$,
\beqarr
     {\sigma_{\rm dc}(0)}/{\sigma_0} &=&
%	({1 - \cos2\pi x })( 2 - {\cos 2\pi x }) / (8\pi^2),  \\
	\left(\frac{1 - \cos2\pi x }{2\pi}\right) \times
           \left(  \frac{2-\cos 2\pi x }{4\pi}\right),  \\
  B  &=&
%	{\cos 2\pi x} / ({ 2 - \cos 2\pi x}).
	\frac{\cos 2\pi x} { 2 - \cos 2\pi x}.
	\label{defB}
\eeqarr
Here we have used the relation $\delta\mu/W = \cot \pi x$
derived from eq.~(\ref{defQPdos}) at $T \ll W$.
{}From eqs.~(\ref{relDCcond}) and (\ref{defrofM}), we have
\bequ
  \frac{ \rho(0) - \rho(M)}{\rho(0)} =
	\frac{ (4 + 3B) M^2 + BM^4}{1+(3+3B)M^2+BM^4},
\eequ
so that %$C = (8-\cos 2\pi x) / ( 2 - \cos 2\pi x)$.
\bequ
   C = \frac{8-\cos 2\pi x} { 2 - \cos 2\pi x}.
\eequ
We see that $C$ monotonically decreases as $x$ is increased;
the maximum is $C=7$ at $x\to0$ and the minimum is $C=3$ at $x=0.5$.

Let us now make a comparison between the above result and
the experimental data in {\LSMO}.
In this case, we should scale the magnetization with
$ M_{\rm tot} / M_{\rm s}$.
Since   $3d$ electrons in {\LSMO} form   localized $S=3/2$ spins
and  mobile electrons,
we have
 $ M_{\rm tot} = \frac32 M + M_{\rm e}$ and $M_{\rm s} = 2$.
Here,
$  M_{\rm e}
  \equiv \frac12( \braket{n_\uparrow} - \braket{n_\downarrow})
  = \frac12(1-x) M$
 is the magnetization of itinerant electrons
in the limit $J/W\to\infty$.
In Fig.~\ref{FigRhoRho0}, $\rho(M_{\rm tot}) / \rho(0)$ at $x=0.175$
is shown as a function of $M_{\rm tot}/M_{\rm s}$.
The experimental data at $x=0.175$  and $T=294{\rm K}$ is taken from
 ref.~\rcite{Tokura94}.
The result at $J/W\to\infty$ shows excellent
agreement with the experimental data of {\LSMO} at
$M_{\rm tot}/M_{\rm s} \simle 0.2$.
It should be noted that no adjustable parameter is used here.

In Fig.~\ref{FigRhoRho0},
the  result at $J/W=4$ and $x=0.175$  is also shown.
Here, $\sigma_{\rm dc}$ and $M_{\rme}$ are obtained numerically.
We see that at $J/W = 4$, the MR value  is already in the
strong coupling limit.
Therefore, together with the previous study,\cite{Furukawa94}\
 we see  that
in the whole range of $J/W$
the MR value is expressed as in eq.~(\ref{defMRgeneric})
at $M \ll 1$.

The experimental data show that
 the coefficient in eq.~(\ref{expUnivFit})
is $C_{\rm exp} \approx 1$ at the heavily
doped metallic system.
As the hole concentration is decreased,
 the value of $C_{\rm exp}$ increases.
In the vicinity of the metal-insulator transition $x\sim 0.15$,
 it approaches the maximum value $C_{\rm exp} \approx 4$.
Enhancement of $C_{\rm exp}$ gives stronger responses to the magnetic field
in the resistivity,
 which is an important feature for applications.
In {\LSMO} at $x=0.175$,
the MR value seems to be already in the strong coupling limit.
Therefore, further increase in $C_{\rm exp}$
can not be expected in this material
by increasing the value of $J/W$ in a chemical way.
The present result shows that $C$ is upper-bounded; at maximum,
the increase of $C$ is slightly less than one order of magnitude.
{}From the point of view of making sensitive MR devices,
it should be better to make attempts in other directions  than increasing
 the value of $C_{\rm exp}$.

In order to obtain
the MR value quantitatively,
it is necessary to calculate the magnetization dependence of
Green's functions in a proper way.
At present, it seems that the infinite-dimensional approach
is the most powerful and successful
 method to investigate the properties of lightly doped {\LSMO}
in the weakly magnetized incoherent-metal state,
 since it can handle thermal
fluctuations in a proper way.
Nevertheless,
we see the discrepancy between
the theoretical result and the experimental data
as the magnetization is increased.
Due to the decrease in the magnitude of the thermal fluctuation
 of spins, quasi-particle excitation recovers its coherence.
Then, in this region,  spatial correlations should also play
important roles.
Further study seems to be necessary to explain the MR value
in the whole region of $M$.

Recently, Millis \etal\cite{Millis9x} have studied
the resistivity of the double-exchange model, emphasizing the role of
the nearest-neighbor correlation of local spins.
They have obtained the increase of the resistivity as
the magnetic moment is increased, which not only fails to
explain the experimental results but also is against the
physical intuition.
The discrepancy seems to be mainly
due to the total neglect of the thermal fluctuation
of local spins and its effect to the self-energies,
which should not be justified near the paramagnetic region.
In contrast to their conclusion, it seems to be clear that
the double-exchange model {\em do} explain
the MR  of the lightly doped {\LSMO}.

To summarize,
double-exchange model in infinite dimension
 with the Lorentzian density of states
is studied as the strong coupling limit of the
Kondo lattice model.
Exact Green's function is obtained analytically.
Resistivity as a function of magnetization is expressed in a simple form.
The MR values in {\LSMO} are well reproduced without any
adjusting parameters.

The author would like to thank Y. Tokura and T. Arima
for fruitful discussions and comments.

\pagebreak

\vspace{2cm}

%
%%%%%%%%%%%%%%%%%%%%%%%%%%%%%%%%%%%%%%%%%%%%%%%%%%%%%%%%%%%%%%%
%%%%%%%%%%%%%%%%%%%%%%%%%%%%%%%%%%%%%%%%%%%%%%%%%%%%%%%%%%%%%%%
%

% GNUPLOT: LaTeX picture
\setlength{\unitlength}{0.240900pt}
\ifx\plotpoint\undefined\newsavebox{\plotpoint}\fi
\sbox{\plotpoint}{\rule[-0.200pt]{0.400pt}{0.400pt}}%
\begin{picture}(1500,900)(0,0)
\font\gnuplot=cmr10 at 10pt
\gnuplot
\sbox{\plotpoint}{\rule[-0.200pt]{0.400pt}{0.400pt}}%
\put(220.0,113.0){\rule[-0.200pt]{292.934pt}{0.400pt}}
\put(220.0,113.0){\rule[-0.200pt]{0.400pt}{184.048pt}}
\put(220.0,113.0){\rule[-0.200pt]{4.818pt}{0.400pt}}
\put(198,113){\makebox(0,0)[r]{0}}
\put(1416.0,113.0){\rule[-0.200pt]{4.818pt}{0.400pt}}
\put(220.0,266.0){\rule[-0.200pt]{4.818pt}{0.400pt}}
\put(198,266){\makebox(0,0)[r]{0.2}}
\put(1416.0,266.0){\rule[-0.200pt]{4.818pt}{0.400pt}}
\put(220.0,419.0){\rule[-0.200pt]{4.818pt}{0.400pt}}
\put(198,419){\makebox(0,0)[r]{0.4}}
\put(1416.0,419.0){\rule[-0.200pt]{4.818pt}{0.400pt}}
\put(220.0,571.0){\rule[-0.200pt]{4.818pt}{0.400pt}}
\put(198,571){\makebox(0,0)[r]{0.6}}
\put(1416.0,571.0){\rule[-0.200pt]{4.818pt}{0.400pt}}
\put(220.0,724.0){\rule[-0.200pt]{4.818pt}{0.400pt}}
\put(198,724){\makebox(0,0)[r]{0.8}}
\put(1416.0,724.0){\rule[-0.200pt]{4.818pt}{0.400pt}}
\put(220.0,877.0){\rule[-0.200pt]{4.818pt}{0.400pt}}
\put(198,877){\makebox(0,0)[r]{1}}
\put(1416.0,877.0){\rule[-0.200pt]{4.818pt}{0.400pt}}
\put(220.0,113.0){\rule[-0.200pt]{0.400pt}{4.818pt}}
\put(220,68){\makebox(0,0){0}}
\put(220.0,857.0){\rule[-0.200pt]{0.400pt}{4.818pt}}
\put(463.0,113.0){\rule[-0.200pt]{0.400pt}{4.818pt}}
\put(463,68){\makebox(0,0){0.2}}
\put(463.0,857.0){\rule[-0.200pt]{0.400pt}{4.818pt}}
\put(706.0,113.0){\rule[-0.200pt]{0.400pt}{4.818pt}}
\put(706,68){\makebox(0,0){0.4}}
\put(706.0,857.0){\rule[-0.200pt]{0.400pt}{4.818pt}}
\put(950.0,113.0){\rule[-0.200pt]{0.400pt}{4.818pt}}
\put(950,68){\makebox(0,0){0.6}}
\put(950.0,857.0){\rule[-0.200pt]{0.400pt}{4.818pt}}
\put(1193.0,113.0){\rule[-0.200pt]{0.400pt}{4.818pt}}
\put(1193,68){\makebox(0,0){0.8}}
\put(1193.0,857.0){\rule[-0.200pt]{0.400pt}{4.818pt}}
\put(1436.0,113.0){\rule[-0.200pt]{0.400pt}{4.818pt}}
\put(1436,68){\makebox(0,0){1}}
\put(1436.0,857.0){\rule[-0.200pt]{0.400pt}{4.818pt}}
\put(220.0,113.0){\rule[-0.200pt]{292.934pt}{0.400pt}}
\put(1436.0,113.0){\rule[-0.200pt]{0.400pt}{184.048pt}}
\put(220.0,877.0){\rule[-0.200pt]{292.934pt}{0.400pt}}
\put(45,855){\makebox(0,0){$\frac{\rho(M_{\rm tot})}{  \rho(0)}$}}
\put(828,23){\makebox(0,0){$M_{\rm tot} / M_{\rm s}$}}
\put(828,824){\makebox(0,0){($x=0.175$)}}
\put(828,824){\makebox(0,0){($x=0.175$)}}
\put(220.0,113.0){\rule[-0.200pt]{0.400pt}{184.048pt}}
\put(1306,812){\makebox(0,0)[r]{Experiment}}
\put(1350,812){\raisebox{-.8pt}{\makebox(0,0){$\Diamond$}}}
\put(220,877){\raisebox{-.8pt}{\makebox(0,0){$\Diamond$}}}
\put(273,868){\raisebox{-.8pt}{\makebox(0,0){$\Diamond$}}}
\put(311,854){\raisebox{-.8pt}{\makebox(0,0){$\Diamond$}}}
\put(349,836){\raisebox{-.8pt}{\makebox(0,0){$\Diamond$}}}
\put(380,809){\raisebox{-.8pt}{\makebox(0,0){$\Diamond$}}}
\put(418,782){\raisebox{-.8pt}{\makebox(0,0){$\Diamond$}}}
\put(448,746){\raisebox{-.8pt}{\makebox(0,0){$\Diamond$}}}
\put(478,719){\raisebox{-.8pt}{\makebox(0,0){$\Diamond$}}}
\put(494,702){\raisebox{-.8pt}{\makebox(0,0){$\Diamond$}}}
\put(532,655){\raisebox{-.8pt}{\makebox(0,0){$\Diamond$}}}
\put(554,619){\raisebox{-.8pt}{\makebox(0,0){$\Diamond$}}}
\put(570,592){\raisebox{-.8pt}{\makebox(0,0){$\Diamond$}}}
\put(592,565){\raisebox{-.8pt}{\makebox(0,0){$\Diamond$}}}
\put(615,542){\raisebox{-.8pt}{\makebox(0,0){$\Diamond$}}}
\put(630,520){\raisebox{-.8pt}{\makebox(0,0){$\Diamond$}}}
\put(646,497){\raisebox{-.8pt}{\makebox(0,0){$\Diamond$}}}
\put(661,475){\raisebox{-.8pt}{\makebox(0,0){$\Diamond$}}}
\put(684,438){\raisebox{-.8pt}{\makebox(0,0){$\Diamond$}}}
\put(706,411){\raisebox{-.8pt}{\makebox(0,0){$\Diamond$}}}
\put(737,375){\raisebox{-.8pt}{\makebox(0,0){$\Diamond$}}}
\put(752,357){\raisebox{-.8pt}{\makebox(0,0){$\Diamond$}}}
\sbox{\plotpoint}{\rule[-0.500pt]{1.000pt}{1.000pt}}%
\put(1306,767){\makebox(0,0)[r]{J/W=4\ \ \,\,}}
\multiput(1328,767)(20.756,0.000){4}{\usebox{\plotpoint}}
\put(1394,767){\usebox{\plotpoint}}
\put(220,877){\usebox{\plotpoint}}
\multiput(220,877)(20.736,-0.902){2}{\usebox{\plotpoint}}
\put(261.19,872.69){\usebox{\plotpoint}}
\put(281.34,867.74){\usebox{\plotpoint}}
\put(301.10,861.44){\usebox{\plotpoint}}
\put(320.19,853.41){\usebox{\plotpoint}}
\put(338.70,844.02){\usebox{\plotpoint}}
\multiput(356,835)(17.385,-11.338){2}{\usebox{\plotpoint}}
\put(391.56,811.26){\usebox{\plotpoint}}
\put(408.22,798.91){\usebox{\plotpoint}}
\multiput(424,786)(16.002,-13.219){2}{\usebox{\plotpoint}}
\put(456.09,759.10){\usebox{\plotpoint}}
\multiput(470,747)(15.014,-14.331){2}{\usebox{\plotpoint}}
\put(501.70,716.73){\usebox{\plotpoint}}
\multiput(515,704)(14.676,-14.676){2}{\usebox{\plotpoint}}
\put(545.83,672.81){\usebox{\plotpoint}}
\multiput(560,658)(14.676,-14.676){2}{\usebox{\plotpoint}}
\multiput(583,635)(14.676,-14.676){2}{\usebox{\plotpoint}}
\put(618.60,598.83){\usebox{\plotpoint}}
\multiput(628,589)(14.361,-14.985){2}{\usebox{\plotpoint}}
\put(661.91,554.09){\usebox{\plotpoint}}
\multiput(674,542)(14.999,-14.347){2}{\usebox{\plotpoint}}
\put(706.43,510.57){\usebox{\plotpoint}}
\multiput(719,498)(14.999,-14.347){2}{\usebox{\plotpoint}}
\put(751.35,467.46){\usebox{\plotpoint}}
\multiput(765,455)(15.014,-14.331){2}{\usebox{\plotpoint}}
\put(797.09,425.23){\usebox{\plotpoint}}
\multiput(810,414)(16.002,-13.219){2}{\usebox{\plotpoint}}
\put(844.60,384.99){\usebox{\plotpoint}}
\multiput(855,376)(16.345,-12.792){2}{\usebox{\plotpoint}}
\put(893.53,346.52){\usebox{\plotpoint}}
\put(910.07,333.99){\usebox{\plotpoint}}
\multiput(923,324)(17.038,-11.853){2}{\usebox{\plotpoint}}
\put(961.00,298.21){\usebox{\plotpoint}}
\put(978.39,286.88){\usebox{\plotpoint}}
\multiput(992,278)(17.511,-11.143){2}{\usebox{\plotpoint}}
\put(1031.03,253.63){\usebox{\plotpoint}}
\put(1048.99,243.22){\usebox{\plotpoint}}
\put(1066.98,232.88){\usebox{\plotpoint}}
\multiput(1082,224)(18.724,-8.955){2}{\usebox{\plotpoint}}
\put(1122.13,204.06){\usebox{\plotpoint}}
\put(1140.64,194.68){\usebox{\plotpoint}}
\put(1159.44,185.89){\usebox{\plotpoint}}
\put(1178.48,177.62){\usebox{\plotpoint}}
\multiput(1196,170)(19.210,-7.859){2}{\usebox{\plotpoint}}
\put(1236.06,153.93){\usebox{\plotpoint}}
\put(1255.38,146.37){\usebox{\plotpoint}}
\put(1274.81,139.07){\usebox{\plotpoint}}
\put(1294.36,132.09){\usebox{\plotpoint}}
\put(1314.03,125.47){\usebox{\plotpoint}}
\multiput(1332,120)(19.856,-6.043){2}{\usebox{\plotpoint}}
\put(1355,113){\usebox{\plotpoint}}
\sbox{\plotpoint}{\rule[-0.400pt]{0.800pt}{0.800pt}}%
\put(1306,722){\makebox(0,0)[r]{J/W=$\infty$\ }}
\put(1328.0,722.0){\rule[-0.400pt]{15.899pt}{0.800pt}}
\put(220,877){\usebox{\plotpoint}}
\put(232,874.34){\rule{3.132pt}{0.800pt}}
\multiput(232.00,875.34)(6.500,-2.000){2}{\rule{1.566pt}{0.800pt}}
\put(245,872.34){\rule{2.891pt}{0.800pt}}
\multiput(245.00,873.34)(6.000,-2.000){2}{\rule{1.445pt}{0.800pt}}
\put(257,869.84){\rule{2.891pt}{0.800pt}}
\multiput(257.00,871.34)(6.000,-3.000){2}{\rule{1.445pt}{0.800pt}}
\put(269,866.84){\rule{2.891pt}{0.800pt}}
\multiput(269.00,868.34)(6.000,-3.000){2}{\rule{1.445pt}{0.800pt}}
\multiput(281.00,865.06)(1.768,-0.560){3}{\rule{2.280pt}{0.135pt}}
\multiput(281.00,865.34)(8.268,-5.000){2}{\rule{1.140pt}{0.800pt}}
\multiput(294.00,860.06)(1.600,-0.560){3}{\rule{2.120pt}{0.135pt}}
\multiput(294.00,860.34)(7.600,-5.000){2}{\rule{1.060pt}{0.800pt}}
\multiput(306.00,855.07)(1.132,-0.536){5}{\rule{1.800pt}{0.129pt}}
\multiput(306.00,855.34)(8.264,-6.000){2}{\rule{0.900pt}{0.800pt}}
\multiput(318.00,849.08)(1.000,-0.526){7}{\rule{1.686pt}{0.127pt}}
\multiput(318.00,849.34)(9.501,-7.000){2}{\rule{0.843pt}{0.800pt}}
\multiput(331.00,842.08)(0.913,-0.526){7}{\rule{1.571pt}{0.127pt}}
\multiput(331.00,842.34)(8.738,-7.000){2}{\rule{0.786pt}{0.800pt}}
\multiput(343.00,835.08)(0.774,-0.520){9}{\rule{1.400pt}{0.125pt}}
\multiput(343.00,835.34)(9.094,-8.000){2}{\rule{0.700pt}{0.800pt}}
\multiput(355.00,827.08)(0.774,-0.520){9}{\rule{1.400pt}{0.125pt}}
\multiput(355.00,827.34)(9.094,-8.000){2}{\rule{0.700pt}{0.800pt}}
\multiput(367.00,819.08)(0.654,-0.514){13}{\rule{1.240pt}{0.124pt}}
\multiput(367.00,819.34)(10.426,-10.000){2}{\rule{0.620pt}{0.800pt}}
\multiput(380.00,809.08)(0.674,-0.516){11}{\rule{1.267pt}{0.124pt}}
\multiput(380.00,809.34)(9.371,-9.000){2}{\rule{0.633pt}{0.800pt}}
\multiput(392.00,800.08)(0.599,-0.514){13}{\rule{1.160pt}{0.124pt}}
\multiput(392.00,800.34)(9.592,-10.000){2}{\rule{0.580pt}{0.800pt}}
\multiput(404.00,790.08)(0.589,-0.512){15}{\rule{1.145pt}{0.123pt}}
\multiput(404.00,790.34)(10.623,-11.000){2}{\rule{0.573pt}{0.800pt}}
\multiput(417.00,779.08)(0.539,-0.512){15}{\rule{1.073pt}{0.123pt}}
\multiput(417.00,779.34)(9.774,-11.000){2}{\rule{0.536pt}{0.800pt}}
\multiput(429.00,768.08)(0.539,-0.512){15}{\rule{1.073pt}{0.123pt}}
\multiput(429.00,768.34)(9.774,-11.000){2}{\rule{0.536pt}{0.800pt}}
\multiput(441.00,757.08)(0.491,-0.511){17}{\rule{1.000pt}{0.123pt}}
\multiput(441.00,757.34)(9.924,-12.000){2}{\rule{0.500pt}{0.800pt}}
\multiput(453.00,745.08)(0.536,-0.511){17}{\rule{1.067pt}{0.123pt}}
\multiput(453.00,745.34)(10.786,-12.000){2}{\rule{0.533pt}{0.800pt}}
\multiput(466.00,733.08)(0.491,-0.511){17}{\rule{1.000pt}{0.123pt}}
\multiput(466.00,733.34)(9.924,-12.000){2}{\rule{0.500pt}{0.800pt}}
\multiput(479.41,718.57)(0.511,-0.536){17}{\rule{0.123pt}{1.067pt}}
\multiput(476.34,720.79)(12.000,-10.786){2}{\rule{0.800pt}{0.533pt}}
\multiput(490.00,708.08)(0.492,-0.509){19}{\rule{1.000pt}{0.123pt}}
\multiput(490.00,708.34)(10.924,-13.000){2}{\rule{0.500pt}{0.800pt}}
\multiput(503.00,695.08)(0.491,-0.511){17}{\rule{1.000pt}{0.123pt}}
\multiput(503.00,695.34)(9.924,-12.000){2}{\rule{0.500pt}{0.800pt}}
\multiput(516.41,680.57)(0.511,-0.536){17}{\rule{0.123pt}{1.067pt}}
\multiput(513.34,682.79)(12.000,-10.786){2}{\rule{0.800pt}{0.533pt}}
\multiput(528.41,667.57)(0.511,-0.536){17}{\rule{0.123pt}{1.067pt}}
\multiput(525.34,669.79)(12.000,-10.786){2}{\rule{0.800pt}{0.533pt}}
\multiput(539.00,657.08)(0.492,-0.509){19}{\rule{1.000pt}{0.123pt}}
\multiput(539.00,657.34)(10.924,-13.000){2}{\rule{0.500pt}{0.800pt}}
\multiput(553.41,641.57)(0.511,-0.536){17}{\rule{0.123pt}{1.067pt}}
\multiput(550.34,643.79)(12.000,-10.786){2}{\rule{0.800pt}{0.533pt}}
\multiput(565.41,628.57)(0.511,-0.536){17}{\rule{0.123pt}{1.067pt}}
\multiput(562.34,630.79)(12.000,-10.786){2}{\rule{0.800pt}{0.533pt}}
\multiput(577.41,615.30)(0.511,-0.581){17}{\rule{0.123pt}{1.133pt}}
\multiput(574.34,617.65)(12.000,-11.648){2}{\rule{0.800pt}{0.567pt}}
\multiput(588.00,604.08)(0.492,-0.509){19}{\rule{1.000pt}{0.123pt}}
\multiput(588.00,604.34)(10.924,-13.000){2}{\rule{0.500pt}{0.800pt}}
\multiput(601.00,591.08)(0.491,-0.511){17}{\rule{1.000pt}{0.123pt}}
\multiput(601.00,591.34)(9.924,-12.000){2}{\rule{0.500pt}{0.800pt}}
\multiput(614.41,576.57)(0.511,-0.536){17}{\rule{0.123pt}{1.067pt}}
\multiput(611.34,578.79)(12.000,-10.786){2}{\rule{0.800pt}{0.533pt}}
\multiput(625.00,566.08)(0.492,-0.509){19}{\rule{1.000pt}{0.123pt}}
\multiput(625.00,566.34)(10.924,-13.000){2}{\rule{0.500pt}{0.800pt}}
\multiput(639.41,550.57)(0.511,-0.536){17}{\rule{0.123pt}{1.067pt}}
\multiput(636.34,552.79)(12.000,-10.786){2}{\rule{0.800pt}{0.533pt}}
\multiput(650.00,540.08)(0.491,-0.511){17}{\rule{1.000pt}{0.123pt}}
\multiput(650.00,540.34)(9.924,-12.000){2}{\rule{0.500pt}{0.800pt}}
\multiput(662.00,528.08)(0.491,-0.511){17}{\rule{1.000pt}{0.123pt}}
\multiput(662.00,528.34)(9.924,-12.000){2}{\rule{0.500pt}{0.800pt}}
\multiput(674.00,516.08)(0.492,-0.509){19}{\rule{1.000pt}{0.123pt}}
\multiput(674.00,516.34)(10.924,-13.000){2}{\rule{0.500pt}{0.800pt}}
\multiput(687.00,503.08)(0.539,-0.512){15}{\rule{1.073pt}{0.123pt}}
\multiput(687.00,503.34)(9.774,-11.000){2}{\rule{0.536pt}{0.800pt}}
\multiput(699.00,492.08)(0.491,-0.511){17}{\rule{1.000pt}{0.123pt}}
\multiput(699.00,492.34)(9.924,-12.000){2}{\rule{0.500pt}{0.800pt}}
\multiput(711.00,480.08)(0.536,-0.511){17}{\rule{1.067pt}{0.123pt}}
\multiput(711.00,480.34)(10.786,-12.000){2}{\rule{0.533pt}{0.800pt}}
\multiput(724.00,468.08)(0.539,-0.512){15}{\rule{1.073pt}{0.123pt}}
\multiput(724.00,468.34)(9.774,-11.000){2}{\rule{0.536pt}{0.800pt}}
\multiput(736.00,457.08)(0.491,-0.511){17}{\rule{1.000pt}{0.123pt}}
\multiput(736.00,457.34)(9.924,-12.000){2}{\rule{0.500pt}{0.800pt}}
\multiput(748.00,445.08)(0.539,-0.512){15}{\rule{1.073pt}{0.123pt}}
\multiput(748.00,445.34)(9.774,-11.000){2}{\rule{0.536pt}{0.800pt}}
\multiput(760.00,434.08)(0.654,-0.514){13}{\rule{1.240pt}{0.124pt}}
\multiput(760.00,434.34)(10.426,-10.000){2}{\rule{0.620pt}{0.800pt}}
\multiput(773.00,424.08)(0.539,-0.512){15}{\rule{1.073pt}{0.123pt}}
\multiput(773.00,424.34)(9.774,-11.000){2}{\rule{0.536pt}{0.800pt}}
\multiput(785.00,413.08)(0.599,-0.514){13}{\rule{1.160pt}{0.124pt}}
\multiput(785.00,413.34)(9.592,-10.000){2}{\rule{0.580pt}{0.800pt}}
\multiput(797.00,403.08)(0.589,-0.512){15}{\rule{1.145pt}{0.123pt}}
\multiput(797.00,403.34)(10.623,-11.000){2}{\rule{0.573pt}{0.800pt}}
\multiput(810.00,392.08)(0.599,-0.514){13}{\rule{1.160pt}{0.124pt}}
\multiput(810.00,392.34)(9.592,-10.000){2}{\rule{0.580pt}{0.800pt}}
\multiput(822.00,382.08)(0.674,-0.516){11}{\rule{1.267pt}{0.124pt}}
\multiput(822.00,382.34)(9.371,-9.000){2}{\rule{0.633pt}{0.800pt}}
\multiput(834.00,373.08)(0.599,-0.514){13}{\rule{1.160pt}{0.124pt}}
\multiput(834.00,373.34)(9.592,-10.000){2}{\rule{0.580pt}{0.800pt}}
\multiput(846.00,363.08)(0.737,-0.516){11}{\rule{1.356pt}{0.124pt}}
\multiput(846.00,363.34)(10.186,-9.000){2}{\rule{0.678pt}{0.800pt}}
\multiput(859.00,354.08)(0.674,-0.516){11}{\rule{1.267pt}{0.124pt}}
\multiput(859.00,354.34)(9.371,-9.000){2}{\rule{0.633pt}{0.800pt}}
\multiput(871.00,345.08)(0.674,-0.516){11}{\rule{1.267pt}{0.124pt}}
\multiput(871.00,345.34)(9.371,-9.000){2}{\rule{0.633pt}{0.800pt}}
\multiput(883.00,336.08)(0.737,-0.516){11}{\rule{1.356pt}{0.124pt}}
\multiput(883.00,336.34)(10.186,-9.000){2}{\rule{0.678pt}{0.800pt}}
\multiput(896.00,327.08)(0.674,-0.516){11}{\rule{1.267pt}{0.124pt}}
\multiput(896.00,327.34)(9.371,-9.000){2}{\rule{0.633pt}{0.800pt}}
\multiput(908.00,318.08)(0.774,-0.520){9}{\rule{1.400pt}{0.125pt}}
\multiput(908.00,318.34)(9.094,-8.000){2}{\rule{0.700pt}{0.800pt}}
\multiput(920.00,310.08)(0.774,-0.520){9}{\rule{1.400pt}{0.125pt}}
\multiput(920.00,310.34)(9.094,-8.000){2}{\rule{0.700pt}{0.800pt}}
\multiput(932.00,302.08)(0.847,-0.520){9}{\rule{1.500pt}{0.125pt}}
\multiput(932.00,302.34)(9.887,-8.000){2}{\rule{0.750pt}{0.800pt}}
\multiput(945.00,294.08)(0.774,-0.520){9}{\rule{1.400pt}{0.125pt}}
\multiput(945.00,294.34)(9.094,-8.000){2}{\rule{0.700pt}{0.800pt}}
\multiput(957.00,286.08)(0.913,-0.526){7}{\rule{1.571pt}{0.127pt}}
\multiput(957.00,286.34)(8.738,-7.000){2}{\rule{0.786pt}{0.800pt}}
\multiput(969.00,279.08)(0.847,-0.520){9}{\rule{1.500pt}{0.125pt}}
\multiput(969.00,279.34)(9.887,-8.000){2}{\rule{0.750pt}{0.800pt}}
\multiput(982.00,271.08)(0.913,-0.526){7}{\rule{1.571pt}{0.127pt}}
\multiput(982.00,271.34)(8.738,-7.000){2}{\rule{0.786pt}{0.800pt}}
\multiput(994.00,264.08)(0.913,-0.526){7}{\rule{1.571pt}{0.127pt}}
\multiput(994.00,264.34)(8.738,-7.000){2}{\rule{0.786pt}{0.800pt}}
\multiput(1006.00,257.08)(0.913,-0.526){7}{\rule{1.571pt}{0.127pt}}
\multiput(1006.00,257.34)(8.738,-7.000){2}{\rule{0.786pt}{0.800pt}}
\multiput(1018.00,250.07)(1.244,-0.536){5}{\rule{1.933pt}{0.129pt}}
\multiput(1018.00,250.34)(8.987,-6.000){2}{\rule{0.967pt}{0.800pt}}
\multiput(1031.00,244.08)(0.913,-0.526){7}{\rule{1.571pt}{0.127pt}}
\multiput(1031.00,244.34)(8.738,-7.000){2}{\rule{0.786pt}{0.800pt}}
\multiput(1043.00,237.07)(1.132,-0.536){5}{\rule{1.800pt}{0.129pt}}
\multiput(1043.00,237.34)(8.264,-6.000){2}{\rule{0.900pt}{0.800pt}}
\multiput(1055.00,231.07)(1.244,-0.536){5}{\rule{1.933pt}{0.129pt}}
\multiput(1055.00,231.34)(8.987,-6.000){2}{\rule{0.967pt}{0.800pt}}
\multiput(1068.00,225.07)(1.132,-0.536){5}{\rule{1.800pt}{0.129pt}}
\multiput(1068.00,225.34)(8.264,-6.000){2}{\rule{0.900pt}{0.800pt}}
\multiput(1080.00,219.07)(1.132,-0.536){5}{\rule{1.800pt}{0.129pt}}
\multiput(1080.00,219.34)(8.264,-6.000){2}{\rule{0.900pt}{0.800pt}}
\multiput(1092.00,213.07)(1.132,-0.536){5}{\rule{1.800pt}{0.129pt}}
\multiput(1092.00,213.34)(8.264,-6.000){2}{\rule{0.900pt}{0.800pt}}
\multiput(1104.00,207.06)(1.768,-0.560){3}{\rule{2.280pt}{0.135pt}}
\multiput(1104.00,207.34)(8.268,-5.000){2}{\rule{1.140pt}{0.800pt}}
\multiput(1117.00,202.07)(1.132,-0.536){5}{\rule{1.800pt}{0.129pt}}
\multiput(1117.00,202.34)(8.264,-6.000){2}{\rule{0.900pt}{0.800pt}}
\multiput(1129.00,196.06)(1.600,-0.560){3}{\rule{2.120pt}{0.135pt}}
\multiput(1129.00,196.34)(7.600,-5.000){2}{\rule{1.060pt}{0.800pt}}
\multiput(1141.00,191.06)(1.600,-0.560){3}{\rule{2.120pt}{0.135pt}}
\multiput(1141.00,191.34)(7.600,-5.000){2}{\rule{1.060pt}{0.800pt}}
\multiput(1153.00,186.06)(1.768,-0.560){3}{\rule{2.280pt}{0.135pt}}
\multiput(1153.00,186.34)(8.268,-5.000){2}{\rule{1.140pt}{0.800pt}}
\multiput(1166.00,181.06)(1.600,-0.560){3}{\rule{2.120pt}{0.135pt}}
\multiput(1166.00,181.34)(7.600,-5.000){2}{\rule{1.060pt}{0.800pt}}
\multiput(1178.00,176.06)(1.600,-0.560){3}{\rule{2.120pt}{0.135pt}}
\multiput(1178.00,176.34)(7.600,-5.000){2}{\rule{1.060pt}{0.800pt}}
\put(1190,169.34){\rule{2.800pt}{0.800pt}}
\multiput(1190.00,171.34)(7.188,-4.000){2}{\rule{1.400pt}{0.800pt}}
\multiput(1203.00,167.06)(1.600,-0.560){3}{\rule{2.120pt}{0.135pt}}
\multiput(1203.00,167.34)(7.600,-5.000){2}{\rule{1.060pt}{0.800pt}}
\put(1215,160.34){\rule{2.600pt}{0.800pt}}
\multiput(1215.00,162.34)(6.604,-4.000){2}{\rule{1.300pt}{0.800pt}}
\put(1227,156.34){\rule{2.600pt}{0.800pt}}
\multiput(1227.00,158.34)(6.604,-4.000){2}{\rule{1.300pt}{0.800pt}}
\multiput(1239.00,154.06)(1.768,-0.560){3}{\rule{2.280pt}{0.135pt}}
\multiput(1239.00,154.34)(8.268,-5.000){2}{\rule{1.140pt}{0.800pt}}
\put(1252,147.34){\rule{2.600pt}{0.800pt}}
\multiput(1252.00,149.34)(6.604,-4.000){2}{\rule{1.300pt}{0.800pt}}
\put(1264,143.34){\rule{2.600pt}{0.800pt}}
\multiput(1264.00,145.34)(6.604,-4.000){2}{\rule{1.300pt}{0.800pt}}
\put(1276,139.34){\rule{2.800pt}{0.800pt}}
\multiput(1276.00,141.34)(7.188,-4.000){2}{\rule{1.400pt}{0.800pt}}
\put(1289,135.84){\rule{2.891pt}{0.800pt}}
\multiput(1289.00,137.34)(6.000,-3.000){2}{\rule{1.445pt}{0.800pt}}
\put(1301,132.34){\rule{2.600pt}{0.800pt}}
\multiput(1301.00,134.34)(6.604,-4.000){2}{\rule{1.300pt}{0.800pt}}
\put(1313,128.34){\rule{2.600pt}{0.800pt}}
\multiput(1313.00,130.34)(6.604,-4.000){2}{\rule{1.300pt}{0.800pt}}
\put(1325,124.84){\rule{3.132pt}{0.800pt}}
\multiput(1325.00,126.34)(6.500,-3.000){2}{\rule{1.566pt}{0.800pt}}
\put(1338,121.84){\rule{2.891pt}{0.800pt}}
\multiput(1338.00,123.34)(6.000,-3.000){2}{\rule{1.445pt}{0.800pt}}
\put(1350,118.34){\rule{2.600pt}{0.800pt}}
\multiput(1350.00,120.34)(6.604,-4.000){2}{\rule{1.300pt}{0.800pt}}
\put(1362,114.84){\rule{3.132pt}{0.800pt}}
\multiput(1362.00,116.34)(6.500,-3.000){2}{\rule{1.566pt}{0.800pt}}
\put(1375,112.34){\rule{1.927pt}{0.800pt}}
\multiput(1375.00,113.34)(4.000,-2.000){2}{\rule{0.964pt}{0.800pt}}
\put(220.0,877.0){\rule[-0.400pt]{2.891pt}{0.800pt}}
\end{picture}

%
%%%%%%%%%%%%%%%%%%%%%%%%%%%%%%%%%%%%%%%%%%%%%%%%%%%%%%%%%%%%%%%
%%%%%%%%%%%%%%%%%%%%%%%%%%%%%%%%%%%%%%%%%%%%%%%%%%%%%%%%%%%%%%%
%

\begin{figcaption}

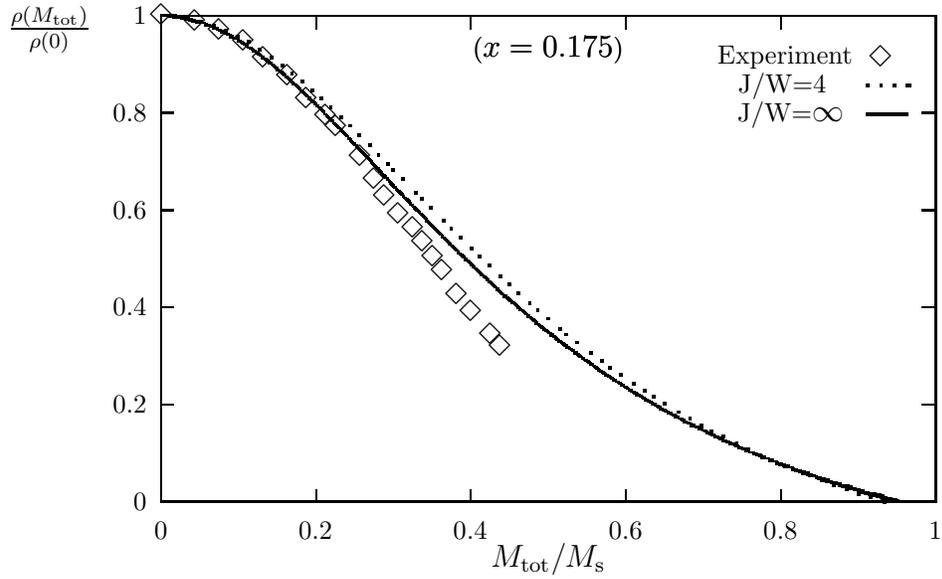
\item Total magnetization dependence of the resistivity
$\rho(M_{\rm tot}) / \rho(0)$
 at $x=0.175$.
The solid (dotted) curve show the theoretical result
at $J/W=\infty$ ($J/W=4$).
The dots in the figure shows the experimental data.
\label{FigRhoRho0}

\end{figcaption}

\pagebreak

%
%
%
%

%\pagebreak
%\newpage
%\input figs

\end{document}